\documentclass[preprint,12pt]{elsarticle}
\usepackage{amssymb}
\usepackage{geometry}
\newgeometry{left=2cm,right=2cm,top=2cm,bottom=2cm}
\usepackage{hyperref}
\hypersetup{colorlinks, citecolor=blue, linkcolor=blue, backref=true}
\usepackage{amsmath}
\usepackage{xcolor}

\usepackage{adjustbox}
\usepackage{multirow}
\usepackage{caption}
\usepackage{subcaption}
\usepackage{algorithm,algorithmic}
\usepackage{caption}
\newlength\myindent
\journal{Healthcare Analytics}

\begin{document}

\begin{frontmatter}

\title{Optimizing Sepsis Care through Heuristics Methods in Process Mining: A Trajectory Analysis}

\author[inst1]{Alireza Bakhshi}
\author[inst1]{Erfan Hassannayebi\corref{mycorrespondingauthor}}\cortext[mycorrespondingauthor]{Corresponding author}\ead{hassannayebi@sharif.edu}
\author[inst2]{Amir Hossein Sadeghi}

\affiliation[inst1]{organization={Department of Industrial Engineering},
            addressline={Sharif University}, 
            city={Tehran},
            country={Iran}}
\affiliation[inst2]{organization={Department of Industrial and Systems Engineering},
            addressline={North Carolina State University}, 
            city={Raleigh},
            state={NC},
            country={USA}}

\begin{abstract}
\textcolor{blue}{Process mining can help acquire insightful knowledge and heighten the system’s performance. In this study, we surveyed the trajectories of 1050 sepsis patients in a regional hospital in the Netherlands from the registration to the discharge phase. Based on this real-world case study, the event log comprises events and activities related to the emergency ward, admission to hospital wards, and discharge enriched with data from lab experiments and triage checklists. At first, we aim to discover this process through Heuristics Miner (HM) and Inductive Miner (IM) methods. Then, we analyze a systematic process model based on organizational information and knowledge. Besides, we address conformance checking given medical guidelines for these patients and monitor the related flows on the systematic process model. The results show that HM and IM are inadequate in identifying the relevant process. However, using a systematic process model based on expert knowledge and organizational information resulted in an average fitness of 97.8\%, a simplicity of 77.7\%, and a generalization of 80.21\%. The analyses demonstrate that process mining can shed light on the patient flow in the hospital and inspect the day-to-day clinical performance versus medical guidelines. Also, the process models obtained by the HM and IM methods cannot provide a concrete comprehension of the process structure for stakeholders compared to the systematic process model. The implications of our findings include the potential for process mining to improve the quality of healthcare services, optimize resource allocation, and reduce costs. Our study also highlights the importance of considering expert knowledge and organizational information in developing effective process models.}
\end{abstract}

\begin{highlights}
\item Surveyed trajectories of 1050 sepsis patients in a regional hospital in the Netherlands.
\item Considered events related to the emergency ward, admission to hospital wards, and discharge.
\item Heuristics and Inductive miners were inadequate in identifying the relevant process.
\item Developed a systematic process model based on expert knowledge.
\item Achieved a fitness of 97.8\%, and a generalization accuracy of 80.21\%.
\end{highlights}


\end{frontmatter}


\section{Introduction}
Process intelligence, a subfield of data science, artificial intelligence, and business process management, is an applicable approach that deals with the practice of collecting and analyzing process data to recognize bottlenecks and improve operational efficiency \cite{arias2022analysis}. Process mining, as an enterprise-wide solution and the main class of process intelligence, is a family of techniques concerning the fields of data science and process management to reinforce the analysis of operational processes based on event logs \cite{van2022process}. The purpose of process mining is to convert event data into insights and steps and also provide solutions to specialists for modeling, documenting, and such cases. Process mining presents three fundamental capabilities: business process discovery, business process conformance checking, and the enhancement of process. During the discovery stage, stakeholders strive to comprehend a process as it is. Process mining traces a process from first to end \cite{yadav2022process}. Therefore, this information is used to attain a process model and assist in apprehending the process execution. In the conformance checking, analysts can compare the existing process model with an event log from the same process to recognize whether the real execution (i.e., as-is process) complies with the expected process execution in the process model (i.e., to-be process) \cite{carmona2018conformance}. Finally, in the enhancement phase, stakeholders are able to improve an existing model through data accumulated in event logs \cite{van2016process}.

The growth of process mining applications is noteworthy in the service industries, especially in healthcare systems. Besides supporting data-driven management and the improvement of healthcare processes, process mining is capable of supporting the resilience of the healthcare system by purveying exhaustive insights into how processes are being accomplished \cite{rojas2016process, guzzo2022process}. Indeed, process mining allows healthcare specialists to comprehend/analyze respective processes so that possible shortcomings are identified and the overall efficiency of the process is improved. Quintano Neira et al. (2019) discovered two bottlenecks and meaningful differences in the performance of some cases through process mining on sepsis patients in a Brazilian hospital  \cite{quintano2019analysis}. Gurgen Erdogan and Tarhan (2018) identified bottlenecks and deviations that were critical for specifying measures in the surgery process \cite{gurgen2018goal}. Motivated by the fact that stakeholders of healthcare processes are looking forward to improving these processes by identifying the problems in the process, and the major studies that have been conducted in this context (e.g., \cite{hendricks2019process,gatta2019clinical,fernandez2021interactive}), we aim to use process mining in the healthcare context in this study. We utilized advanced applied machine learning in process mining techniques to examine the journeys of 1050 patients suspected of having sepsis who were admitted to the emergency room of a hospital in the Netherlands. In general, the trajectory of a patient (no matter the kind of disease) from the entrance to the emergency room until his discharge can be considered a process. According to this, we investigate such a process for patients with suspected sepsis. The objectives of the process mining in this study are as follows: (1) Discovering the relevant process and acquiring insights into the patient trajectories, (2) Validating adherence with medical guidelines for sepsis treatment, and (3) Assessing the use of process mining approaches in the discussed context. This study also focuses on the following research questions:
\begin{itemize}
    \item Are both Inductive Miner (IM) and Heuristics Miner (HM) suitable for detecting the trajectory of sepsis patients? 
    \item Are the following medical guidelines for treating sepsis patients followed accurately?
    \begin{itemize}
        \item Patients should be administered antibiotics within one hour.
        \item Lactic Acid measurements should be performed within three hours.
    \end{itemize}
    \item What are the pathways of patients that involve either being discharged without admission, being admitted to regular care, being admitted to intensive care, or being admitted to regular care and then transferred to intensive care, depending on the manner in which they were admitted to the emergency room?
    \item What is the trajectory of patients who returned within 28 days?
\end{itemize}
	
To elaborate on the research questions, we would like to discover the process of the trajectory of sepsis patients through two recognized algorithms (i.e., IM and HM) and then evaluate and compare the performance and efficiency of each in detecting such a process. In addition, we make a comparison between the process models obtained by these two methods and a systematic process model developed based on organizational information to reveal the effectiveness of this model compared to the previous two. After discovering the respective process model, we seek to analyze and compare the medical guidelines with what actually happened by the conformance checking techniques to identify possible defects and any discrepancies with the medical guidelines. Finally, we endeavor to identify the trajectory of problematic patients, e.g., those who returned to the hospital within 28 days, and those whose condition deteriorates and are sent to the intensive treatment ward after normal treatment.

Note that most similar studies focused on discovering the selected process through different tools such as Process Mining frameworks (ProM). Some studies identified the process through well-known process discovery algorithms as well \cite{wafda2022systematic}. In fact, these studies that addressed process discovery have ignored the process structure because these tools and algorithms cannot intrinsically consider the process structure and merely identify the process based on the related event log. In this case, the obtained process model is not usually tangible for the process stakeholders. In addition, conforming the event log with the discovered process to answer possible questions from such models may not accurately reflect the results. Therefore, if a model with principles and rules is created through the organizational structure and knowledge, it will bring more insight to the stakeholders, and the conformance-checking phase will be accomplished precisely. However, to the best of our knowledge, the related process mining papers that address such topics in healthcare systems remain scarce. The main contribution of our study is that we intend to bridge this gap in research specifically by presenting IM and HM process models for a process related to the trajectory of sepsis patients and comparing them with a systematic process model and then addressing the conformance-checking phase from the systematic process model.

The rest of this paper is organized as follows: Section~\ref{sec:lit} addresses the literature review on different approaches used for process discovery related to healthcare systems and recounts the works that researchers have taken in these studies. In Section~\ref{sec:method}, the methodology and algorithms used for process mining are explained in detail. Section~\ref{sec:case} represents explanations about the event log, required actions for data pre-processing, and variant analysis. In Section~\ref{sec:result}, we provide results obtained by process discovery and conformance checking as well as endeavoring to answer research questions. Managerial implications are presented in Section~\ref{sec:discussoin}. Finally, we conclude the paper in Section~\ref{sec:conclusion}.

\section{Literature review} \label{sec:lit}
This section provides recent background on process mining in healthcare systems. We try to review different topics and compare them with our work.

First, the classification of processes in the context of healthcare is discussed. In general, \textcolor{blue}{hospital processes are classified into medical treatment processes and organizational processes \cite{kraus2021digital}}. Medical treatment involves overseeing the care of patients from diagnosis to the implementation of measures aimed at alleviating their symptoms or illnesses. This includes various diagnostic tests, procedures, medications, therapy, \textcolor{blue}{and follow-up care to evaluate the success of the treatment and ensure the patient's health and well-being \cite{salameh2020preferred,klein2020treatment}}. In return, organizational processes concentrate on the organizational knowledge of processes, receiving joint information from specialists \cite{rebuge2012business}. Another classiﬁcation divides operational healthcare processes into two classes: the ﬁrst one correlates to nonselective care, consisting of medical emergencies; and the second one associates with selective care, including routine and non-routine methods \cite{mans2015process}. The real-life log we reviewed is about the process of the sepsis patient's trajectories from the time of registration to discharge; therefore, it is placed in the field of medical treatment processes. Note that one of the developed process models, the systematic process model, which will be explained in the next sections, is based on organizational knowledge; hence, our work is related to organizational processes as well.

Now, we make a comparison of the types of data. Data defines processes that can be analyzed. \textcolor{blue}{Thus, they can be collected to create event logs, letting the process mining approaches be performed \cite{de2022process}. These event logs are concerned with the information documents, including events and activities, accomplished through a process within a respective case. The first category is based on data used in case studies that can include different types of crucial symptoms, events, the patient's data, and doses of infusion or inhalation medicines \cite{van2012process,noshad2022signal}}. The second category of data is related to administrative systems, clinical support systems, healthcare logistic systems, and medical devices \cite{mans2013process,munoz2022process}. The event log data analyzed in this research is of the first category.

\textcolor{blue}{There are a series of tools that allow process mining approaches to be used to develop process models for analysis. ProM is one of these tools that executed extensive methods and algorithms; hence, a large number of studies have used it for process mining (e.g., \cite{revina2023approach,de2022process,cho2014systematic,helm2015first})}. Other tools have been used in different papers as well. For example, Disco contains a licensed tool with a visual interface for process models and multiple filtering options in event logs. RapidProM includes a workflow description functionality based on the plugins and the data analysis solutions executed in ProM and RapidMiner, respectively. \textcolor{blue}{Some studies related to process mining have used Disco (e.g., \cite{partington2015process, miclo2015rtls,habibagahi2022co}) and others have used RapidProM (e.g., \cite{mans2015process})}. We also used both ProM and Disco in this research (ProM for conformance checking and Disco for variant analysis).

Besides, so far, there have been numerous process discovery algorithms in the literature. The prevalent algorithms operated in case studies are HM and Fuzzy Miner. HM is a discovery algorithm that can develop process models and performs well in dealing with noisy data \cite{mans2015process,ganesha2017analyzing,neamsirorat2015analysis}. Fuzzy Miner is a configurable discovery algorithm that creates several models at various detailed levels, assisting to deal with unstructured processes \cite{forsberg2016analyzing, jaisook2015time}. IM and Alpha Miner are also other known algorithms in process mining. IM seeks to remove infrequent activities and routes \cite{mannhardt2017analyzing, ganesha2017best}. Alpha miner is the most uncomplicated process mining algorithm, that forms Petri nets from the detected model \cite{zhou2014process, delias2014applying}. In this study, since HM deals better with anomalies, we used it. In addition, despite the effectiveness of IM in determining specific arcs and identifying deviations, it has been very little used in previous health-oriented studies; thus, we also decided to use it. Then, we compare the process models obtained by them with the systematic process model based on organizational knowledge.

Finally, regarding the different types of process mining activities carried out in previous similar papers, it can be summarized these things to process discovery, conformance checking, variant analysis, process enhancement, performance analysis, and predictive monitoring. There are abundant studies on process discovery, indicating the importance of understanding a process. For instance, \cite{ganesha2017best} intended to discover a process model for the utilization of physical resources in hospitals. \cite{neamsirorat2015analysis} sought to discover the process in the surgical department of a hospital. Many studies have also focused on conformance checking (e.g., \cite{forsberg2016analyzing}). \cite{fernandez2016interactive} addressed interactive pattern recognition in cardiovascular disease management. \cite{caron2014process} concentrated on a process mining-based analysis of unfavorable events in care processes. Some studies also explored both process discovery and conformance checking (e.g., \cite{mannhardt2017analyzing,ganesha2017analyzing}. Ref. \cite{delias2014applying} addressed variant analysis in an emergency department in detail. Since healthcare systems are of outstanding importance in every area, studies have also focused on improving health-oriented processes (e.g., \cite{zhou2014process,fakhrabad2022evaluating,fakhrabad2023impact}). Ref. \cite{rojas2016process, jaisook2015time} addressed performance analysis in the context of healthcare systems. Ref. \cite{di2016clustering} sought to monitor clustering-based predictive processes. \textcolor{blue}{Another study used process mining to detect differences in the health services utilization pattern of patients during the COVID-19 pandemic and mandatory lock-downs in 2020 compared to the prior four years \cite{augusto2022process}. A more recent study proposes a new process complexity measure based on graph entropy and finds that various process complexity measures have a correlation with the quality of the discovered process models \cite{augusto2022connection}. Note that we did not restrict the scope of our study to one of these activities; hence, we address three activities from the aforementioned process mining activities, i.e., process discovery, variant analysis, and conformance checking. In fact, since the real-life log related to sepsis patients in a hospital has been studied, it is essential to analyze the process, its key variants, and compliance with medical guidelines at the same time.}

As mentioned throughout this section, most of the similar papers have mainly focused on one of the process mining activities, i.e., merely process discovery, or others. In general, among the mentioned activities, most of the studies have either dealt with the recognition of processes or conformance checking, or performance analysis. Although some studies examined a series of these activities, there is still a lack of such studies, especially in real-life logs related to emergency wards in hospitals. Therefore, in this research, not only the identification of the emergency-oriented process is accomplished, but also the variant analysis and the conformance checking are scrutinized, e.g., analyzing the difference between the medical instructions and what is actually being executed. Moreover, for process discovery, the attention of researchers is mainly drawn to HM or fuzzy miner algorithms, because they have a significant ability to cope with noisy data. However, investigating the performance of other effective algorithms, e.g., IM that has the ability to deal with infrequent behavior has often been neglected in this context. Also, the process models obtained by these algorithms may not be explicit to stakeholders; hence, it is better to create a process model based on organizational knowledge to tackle this issue as well. Notwithstanding this has rarely been considered in other similar studies, we endeavor to fill this gap by comparing two process models acquired by HM and IM to a process model based on organizational knowledge.

\section{Methodology} \label{sec:method}
In this study, given that we are aware of the widespread importance of process mining in the healthcare context and based on the research questions, we are interested in analyzing the process concerned with the sepsis patients of a hospital in the Netherlands from their registration in the emergency ward to their discharge. One of the reasons for selecting this group of patients is to limit the problem space for a more detailed study. The event log consists of information about the emergency ward and the hospital laboratory. Indeed, process mining can help stakeholders to take a closer look at the trajectory of patients in the hospital and identify/improve the possible system defects \cite{yadav2022process}. In this regard, after a primary understanding of the event log, variant analysis by Disco \cite{porouhan2022improving}, and pre-processing phase, we aim to address the respective process discovery through two recognized methods, i.e., IM and HM to see which one is suitable for detecting this process. Note that the structure of these two algorithms is dissimilar from each other. In the next step, we compare the two obtained process models with a systematic model created based on organizational information. Since the structure of these models is different, we try to compare the three exhaustively and mention the best one in identifying the relevant process. Afterward, we seek to check the conformance of the cases mentioned before. Namely, we aim to analyze the extent to which the actual process conforms to the target process. For this purpose, in addition to PM4PY documentation, we use the multi-perspective conformance-checking technique as well. Multi-perspective conformance checking evaluates a process model's compliance from multiple angles at the same time. This means that deviations may appear in a different perspective compared to the control viewpoint, or deviations from the control viewpoint can be explained based on other perspectives. We used the ProM Plug-in-Multi-Perspective Process Explorer (MPE), a tool designed to explore a multi-perspective process to improve and analyze compliance \cite{erdogan2022multi}. Note that MPE performance is merely not abridged to what has been mentioned, not only does it integrate existing decision-mining methods into a scalable and flexible tool but it also provides interactive information visualizations and filtering features. In general, MPE provides three main features:

\begin{itemize}
    \item Combining multiple approaches for conformance checking, identifying deviations, and analyzing performance to produce a comprehensive evaluation of a system.
    \item Finding decision-making rules that govern the process through effective interaction.
    \item Analyzing event logs internally by filtering and exploring based on context-sensitive graphs and monitoring probe types.
\end{itemize}

Keep in mind that users, as well as event information, are important to this technique. Finally, we attempt to find the decision rules based on the features available in the event log regarding the trajectories patients have taken. In what follows, we explain each of the process mining algorithms in detail.

As mentioned, we aim to survey the effectiveness and performance of IM and HM to identify the trajectory of sepsis patients and illustrate the interactions within the process well. First, we explain IM and elucidate how it works, then we interpret the mechanism of HM similarly.

\subsection{Inductive Miner}
\begin{figure}[H]
\centering
\includegraphics[width=10cm]{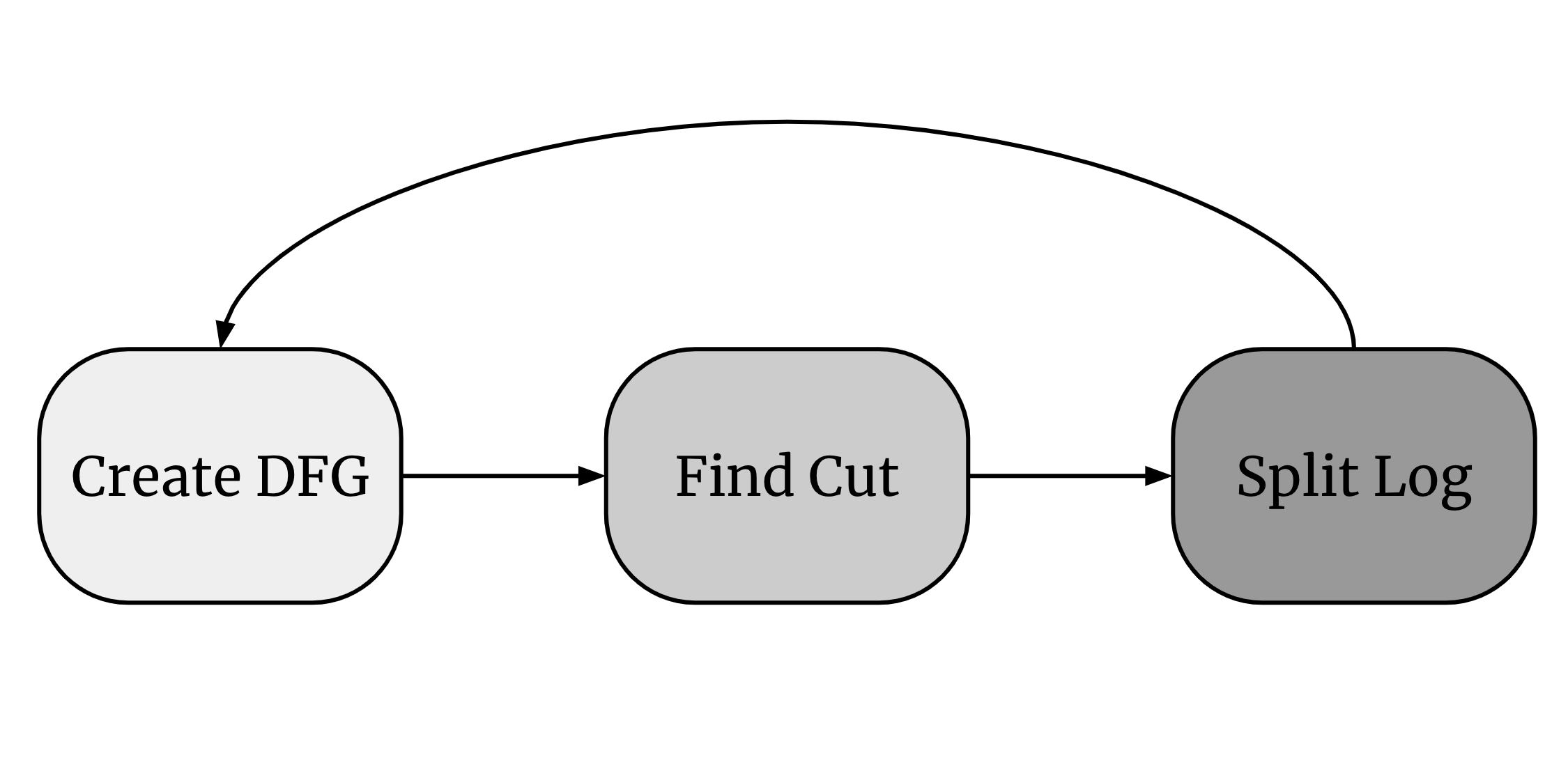}
\caption{Three steps of IM method.}
\label{fig:IM}
\end{figure}

IM leans on creating a directly-follows graph (DFG) from the event log and then taking advantage of DFG to discover different process relations through the detection of various cuts \cite{hachicha2021using,leemans2015scalable}. To clarify, the main concept of IM has originated from the methodology of detecting different divisions of the arcs in DFG and using the smaller segments to depict the implementation succession of activities \cite{cao2023discovery}. IM includes three major stages which are recursively performed. the mechanism of this algorithm is depicted in Figure~\ref{fig:IM}. Note that the only input for IM is an event log ($L$).

Based on Figure~\ref{fig:IM}, the first stage of this algorithm is related to computing DFG. DFG connects, for example, activity $\alpha$ to another activity $\beta$ if activity $\beta$ ensues chronologically right after activity $\alpha$. In the second stage, IM uses DFG to find various cuts, including exclusive choice cuts, sequence cuts, concurrent cuts (parallel cuts), and loop cuts. The exclusive choice cut (XOR-cut) classifies activities in such a way that activities belonging to distinct groups have nothing to do with each other. Then, the length of the connected components in each group is measured. If it is greater than one (i.e., the connected components contain more than one group), an XOR cut will be discovered and the connected components will be returned as a cut partition.

In the sequence cut detection, all pairwise reachable groups are firstly merged, implying that the activities are partitioned into groups $\{P_1,\dots,P_n\}$ so that there are no two activities in different groups that are pairwise reachable. following, we merge the pairwise unreachable nodes. Then, the resulting groups $\{P_1,\dots,P_n\}$ are ordered and if $k>1$, a sequence cut is detected, and $\{P_1,\dots,P_n\}$ are returned as a cut partition. 

In concurrent cut detection, activities are firstly classified in such a way that each group has a start activity and an end activity, and the activities between these groups are of directly-follows relations as well. Then, we explore groups that include no start activity and no end activity. Such groups are merged with another arbitrary group. If the partition comprising the resulting sub-groups has a length greater than one, a concurrent cut will be detected. Thus, the resulting partition is returned as a cut partition. 

For detecting a loop cut, a group $P_1$ is created, including all the start and end activities. Afterward, $\{P_2,\dots,P_n\}$ are created from all the rest activities so that $\alpha \to \beta$ with $\alpha \in P_i$ and $\beta \in P_j$ invariably shows that $i<j$. Therefore, there can be no connection between a group with a smaller index to a group with a bigger index. In the following, every group $P_i$ ($i \in [2,n]$) will be merged that is connected from a start activity or to an end activity with $P_1$. Besides, groups with $P_1$ are also merged that have an activity connected to one but not to all start activities and groups that have an activity that is connected from one but not to all end activities. If $n>1$, a loop-cut is detected, and $\{P_1,\dots,P_n\}$ are returned as a cut partition.

If each of these four types of cuts is successfully detected, the algorithm will return a cut operator and a cut partition. Given the returned cut-operator and cut-partition, IM divides the event log into one or more sub-logs. This stage, which is the last step of IM, is called event log splitting. According to the detected cut, a diﬀerent split function is exerted. Hence, there are four split functions (i.e., exclusive choice split (XOR), sequence split, concurrent split, and Loop split). In an XOR split, for $n$ cut partitions $\{\sigma_1, \dots, \sigma_n \}$, the log $L$ is divided into $n$ sub-logs $\{L_1,\dots,L_n\}$. To construct a sub-log $L_i$  ($i\in[1,n]$), the implementation is iterated over all traces  $\{T_1,\dots,T_n\}$ in $L$. Those traces are added to $L_i$, where each activity in the trace is incorporated in $\sigma_i$. 

In a sequence split, the function takes a cut partition $\{\sigma_1, \dots, \sigma_n \}$ and a log as input. In what follows, for every $\sigma_i$, a sub-log $L_i$ is built, then this is iterated for all the traces $\{t_1,\dots,t_x\}$ in the log. For every trace $t_x$, a new vacant trace $t_x^\prime$ will be created. Afterward, it is iterated over all the activities in $t_x$. When activity in $t_x$ is contained in $\sigma_i$, this activity and the following ones, until they are comprised in $\sigma_i$, will be appended to $t_x^\prime$. Once observing an activity that is not contained in $\sigma_i$, $t_x^\prime$ is appended to $L_i$, a new vacant trace is created, and this cycle will be iterating over the activities in the next trace $t_{x+1}$. 

Based on a concurrent split, the function takes a cut partition $\{\sigma_1, \dots, \sigma_n \}$ and the log as input. Then, the sub-logs $\{L_1,\dots,L_n\}$ are built. To construct a sub-log $L_i$, the implementation is iterated through the traces $\{t_1,\dots,t_x\}$ of the log. For every activity in a trace $t_j$, it should be analyzed whether this activity is included in $\sigma_i$ or not. If so, it is added to the trace $t_j$ in $L_i$. 

Based on a looping split, The function takes a cut partition $\{\sigma_1, \dots, \sigma_n \}$ and the log as input. To create the sub-logs $\{L_1,\dots,L_n\}$, the implementation is started by iterating over the cut partitions. For every cut partition $\sigma_i$, a new sub-log $L_i$ is constructed, and it is iterated over the traces of the log. Then, the sequences of activities $\{\alpha_1,\dots,\alpha_n\}$ where, with $j\in[1,n]$, every $\alpha_j \in \sigma_i$ and $\alpha_0 \notin \sigma_i$ and $\alpha_{n+1} \notin \sigma_i$ will be explored. Finally, all the sequences discovered are added to the current sub-log $L_i$. 

Note that the algorithm is recursively exerted to each sub-log, i.e., DFG is computed for every sub-log, then searches are performed to find the mentioned cuts, and finally, the log is split according to the detected cuts. This cycle continues until a so-called base case is detected. Once a base case is discovered (i.e., checking before computing DFG in each iteration), a branch of the recursion comes to an end. If neither base case nor any cut is discovered, a so-called fall-through (e.g., empty traces, activity once per trace, activity concurrent, and the like) will be applied.

\subsection{Heuristics Miner}
Now, we describe how HM operates. HM searches for the control-flow perspective of a process model \cite{weijters2006process}. This algorithm solely considers the order of the events within a case, namely, the order of events among cases is not important \cite{weijters2006process,winter2020discovering}. The input data to this algorithm is just an event log so it uses the three main features of the log, i.e., case id, activity, and time to find the various types of relationships between activities. Note that to detect a process model on the basis of the event log, the log should be surveyed for finding causal dependencies, e.g., if activity $\alpha$ is always followed by activity $\beta$, there is probably a dependency relation between these activities. By and large, This algorithm consists of three main steps: (1) Creating a dependency graph, (2) Identifying AND/XOR-split/join and non-observable tasks, and (3) Mining long-distance dependencies \cite{weijters2006process}. 

To sum up, in this section, we endeavored to explain the process discovery and conformance-checking techniques used in this study. In the next section, we analyze the event log, how to address data pre-processing, and scrutinize variants in detail.

\section{Case study} \label{sec:case}
In this section, we provide information about the real-life log features and the required pre-processing it needs to be ready for the process discovery phase and the later phase, i.e., conformance checking. Besides, we identify the crucial traces and analyze them to better comprehend the event log and the phases we are going to address in the following sections.

\subsection{Data dictionary }
This study analyzes an event log about the trajectories of patients admitted to the emergency ward in a Dutch hospital collected from their registration span in emergency rooms until their discharge over a one-and-a-half-year period. This hospital has nearly 700 beds at distinct sites recoursed by approximately 50,000 patients in a year. Note that we aim to concentrate on a single group of patients suspected of bearing sepsis (i.e., those for which a distinguishing treatment is to be expected) to prevent a prevalent problem in process mining due to the innate complexity and the healthcare processes' ﬂexibility. To be specific, sepsis is a life-threatening condition typically caused by an infection. Usual symptoms of sepsis are: (1) increased body temperature, (2) increased heart rate, (3) high/heavy respiratory rate, and (4) abnormal white blood cell count. These symptoms are simultaneously known as Systemic Inflammatory Response Syndrome (SIRS) criteria used as signs of sepsis.
In general, the procedure for collecting this event log is summarized in three main steps, which are as follows:
\begin{itemize}
    \item The sepsis patient triage record encompasses the triage time, symptoms (such as SIRS criteria for sepsis), diagnostic tests ordered, and times injections of liquid and antibiotics;
    \item Records available in the laboratory by taking several blood tests from patients; 
    \item Data recorded by ﬁnancial systems about the extra trajectory of patients. 
\end{itemize}

Note that the required documents are stored in the ERP system of the hospital. Accordingly, the event log contains 16 activities for more than 1050 sepsis patients (i.e., 1050 cases), which is recorded for 15215 different events during nearly 1.5 years in the hospital ERP system. The number of variants or traces recognized from these cases is 890. Besides, the identified activities are classified into logistical and medical or treatment activities:
\begin{itemize}
    \item Three activities related to the registration/triage in the emergency rooms (i.e., ER Registration, ER Triage, and ER Sepsis Triage); 
    \item Three activities for measuring CRP, Leukocytes, and Lactic Acid; 
    \item Two activities related to required injections (i.e., IV Antibiotics and IV Liquid);
    \item Two activities related to the transition to normal or intensive care; 
    \item Five activities for different types of hospital discharge (i.e., Release A to E);
    \item An activity related to the return of patients at a subsequent time.
\end{itemize}

Moreover, 28 data attributes have been recorded, e.g., the group responsible for the activity, the patient age, the results of blood measurement attributes (i.e., lactic acid, leukocytes, and CRP), information about the diagnosis of disorders/illegal materials in different organs of the patient's body (e.g., diagnostic sputum, diagnostic liquor, and the like), and patient information from checklists recorded in the triage documents (e.g., infusion, hypotensive, hypoxia, oliguria, and so on). The timestamps of events have been randomized, but the time between events in a trace has not been changed. A brief excerpt for the first two cases from the studied event log is demonstrated in Table~\ref{tab:eventlog}. Since there are diverse features in the event log, key features are mentioned in Table~\ref{tab:eventlog}. Note that the bold \textbf{(.)} values were missing data that were filled in by the criteria we describe in detail in the Data pre-processing section.

\begin{table}[H]
\caption{Excerpt from a sepsis event log}
\label{tab:eventlog}
\begin{adjustbox}{width=\textwidth}
\begin{tabular}{lllllllll}
\hline
Case ID & Activity & Timestamp & CRP & Responsible Group & Leukocytes & Lactic Acid & Age & Infection Suspected \\
\hline
\multirow{22}{*}{A} & ER Registration & 00:15:41 & (\textbf{21}) & A & (\textbf{9.6}) & (\textbf{2.2}) & 85 & True \\
 & Leukocytes & 00:27:00 & (\textbf{21}) & B & 9.6 & (\textbf{2.2}) & 85 & True \\
 & CRP & 00:27:00 & 21 & B & 9.6 & (\textbf{2.2}) & 85 & True \\
 & Lactic Acid & 00:27:00 & 21 & B & 9.6 & 2.2 & 85 & True \\
 & ER Triage & 00:33:37 & 21 & C & 9.6 & 2.2 & 85 & True \\
 & ER Sepsis Triage & 00:34:00 & 21 & A & 9.6 & 2.2 & 85 & True \\
 & IV Liquid & 00:03:47 & 21 & A & 9.6 & 2.2 & 85 & True \\
 & IV Antibiotics & 00:03:47 & 21 & A & 9.6 & 2.2 & 85 & True \\
 & Admission NC & 00:13:19 & 21 & D & 9.6 & 2.2 & 85 & True \\
 & CRP & 00:00:00 & 109 & B & 9.6 & 2.2 & 85 & True \\
 & Leukocytes & 00:00:00 & 109 & B & 8.7 & 2.2 & 85 & True \\
 & Leukocytes & 00:00:00 & 109 & B & 9.6 & 2.2 & 85 & True \\
 & CRP & 00:00:00 & 47 & B & 9.6 & 2.2 & 85 & True \\
 & Leukocytes & 00:00:00 & 47 & B & 10.7 & 2.2 & 85 & True \\
 & CRP & 00:00:00 & 15 & B & 10.7 & 2.2 & 85 & True \\
 & CRP & 00:00:00 & 9 & B & 10.7 & 2.2 & 85 & True \\
 & Leukocytes & 00:00:00 & 9 & B & 13 & 2.2 & 85 & True \\
 & Leukocytes & 00:00:00 & 9 & B & 11.3 & 2.2 & 85 & True \\
 & CRP & 00:00:00 & 9 & B & 11.3 & 2.2 & 85 & True \\
 & CRP & 00:00:00 & 6 & B & 11.3 & 2.2 & 85 & True \\
 & Leukocytes & 00:00:00 & 6 & B & 10.9 & 2.2 & 85 & True \\
 & Release A & 00:15:00 & 6 & E & 10.9 & 2.2 & 85 & True \\
 \hline
\multirow{12}{*}{B} & ER Registration & 00:04:24 & (\textbf{240}) & A & (\textbf{13.8}) & (\textbf{0.8}) & 45 & True \\
 & ER Triage & 00:17:19 & (\textbf{240}) & C & (\textbf{13.8}) & (\textbf{0.8}) & 45 & True \\
 & CRP & 00:36:00 & 240 & B & (\textbf{13.8}) & (\textbf{0.8}) & 45 & True \\
 & Lactic Acid & 00:36:00 & 240 & B & (\textbf{13.8}) & 0.8 & 45 & True \\
 & Leukocytes & 00:36:00 & 240 & B & 13.8 & 0.8 & 45 & True \\
 & ER Sepsis Triage & 00:15:45 & 240 & A & 13.8 & 0.8 & 45 & True \\
 & IV Liquid & 00:33:48 & 240 & A & 13.8 & 0.8 & 45 & True \\
 & IV Antibiotics & 00:33:55 & 240 & A & 13.8 & 0.8 & 45 & True \\
 & Admission NC & 00:17:08 & 240 & F & 13.8 & 0.8 & 45 & True \\
 & CRP & 00:00:00 & 226 & B & 13.8 & 0.8 & 45 & True \\
 & CRP & 00:00:00 & 75 & B & 13.8 & 0.8 & 45 & True \\
 & Release A & 00:00:00 & 75 & E & 13.8 & 0.8 & 45 & True \\
 \hline
\end{tabular}
\end{adjustbox}
\end{table}

\subsection{Data pre-processing }
This section outlines the necessary pre-processing on the event log. The pre-processing of the event log endeavors to find and eliminate noise events, activities, or missing values bringing about unfavorable behavior. This is a prelude to modifying the log, which is essential for the accurate performance of process mining duties. In general, there are almost no key problems in the event log provided by the hospital ERP system. The defect found in this dataset is related to some missing values which could be derived from incomplete data entry, equipment malfunctions, lost files, and many other reasons. These missing values are of missing at random type, implying that the basis for missing values can be clarified by variables on which we have comprehensive information as there is some relationship between the missing data and other values. 

First, in 55 cases, either the information about the first starting activity or the first two starting activities was documented incompletely. These missing values that are all boolean types are merely concerned with the information from checklists recorded in the triage documents (e.g., hypoxia, and the like) and the diagnosis of disturbances/illegal materials in the patient's body (e.g., diagnostic thorax, and the like). Note that in the next activities of such cases, all data of triage documents have been recorded; thus, given that these data have a certain hierarchy, we have also filled these missing values with the same data of the first activity in which these attributes are recorded. 

Second, in cases where the trajectory of patients includes the measurement of CRP, lactic acid, or leukocytes, no data has been recorded for each of them until the first measurement. Once each of which is measured for the first time, their data are recorded and this data will not change until the next measurement of the same indicator. These empty cells, which are mainly for the time when patients are in a registration process (i.e., ER Registration, ER Triage, and ER Sepsis Triage), can be filled through the hierarchical relationships mentioned above. Therefore, data recorded in the first measurement for each indicator can be used to fill the vacancy associated with that index before the first measurement. Meanwhile, in cases where the above indicators have been measured but the number has not been recorded due to any reason, we consider the same number recorded at the time of the next measurement.

\subsection{Variant discovery and analysis}
Analyzing variants can help how to follow the principles and rules in the planned framework of any organization (hereon the hospital). As mentioned, out of 1,050 recorded, there are 845 variants, signifying that all patients nearly underwent dissimilar trajectories. The most frequent variant is 4\% of the patients, which includes 46 cases (e.g., AM). This variant that took less than a day is as follows: \textit{ER Registration→ER Triage→ER Sepsis Triage→CRP→Leukocytes}. Specifically, this is apparently for suspicious patients who have spent the registration procedures and then the measurement of CRP and Leukocytes, which in turn demonstrates that their condition was stable and they did not need any special treatment or hospitalization. In general, about 15\% of patients (i.e., 161 cases) went through the most five common variants, which took an average of one hour to finish.

In addition, in terms of the number of repetitions of events, the longest variant is associated with the NGA case, including 185 activities in 115 days. Of these 185 events, 73 times were related to measuring leukocytes, 68 times to measuring CRP, 30 times to measuring Lactic Acid, and 3 times to admission to normal care. The rest of the activities have also been done once each (e.g., ER registration, IV liquid, and so on). In 844 variants, one of the indicators (i.e., CRP, lactic acid, and leukocytes) is going to be measured, indicating that the measurement of these indicators is one of the vital steps for the diagnosis of sepsis cases. In almost 162 variants, the measurement of all three indicators (i.e., CRP, acid lactic, leukocytes) based on blood tests has been repeated. By observing these variants, it can be concluded that these three key indicators must have regularly been measured for patients who have been infected so that if any of these are out of their normal range, they will immediately be transferred to normal care wards or in more acute cases to intensive care wards. Indeed, these patients should be monitored by specialists for at least two days in the hospital.

By and large, the trajectory of 212 patients corresponds to 20\% of the most common variants, of which 3 patients experienced a 2-day process and 6 of whom were in the one-day process. The path of 163 patients took less than one day, and the rest of these patients were in the hospital for between 3 and 10 days. This indicates that the patients with rather abnormal symptoms have spent at least one day in the hospital for further monitoring. In addition, the longest time spent on variants is 422 days (e.g., HG case). Such a variant belongs to patients who, after spending about 5 days in the hospital and being monitored, returned to the hospital about 417 days later, that is, after about a year, they again had suspicious symptoms of sepsis and went to the hospital. Furthermore, about 3.7\% of patients, i.e., 39 cases in 39 variants, were transferred from the normal care ward to the intensive care ward due to the aggravation of their condition. The fastest time among these 39 variants took two days (e.g., WKA case) and the longest one took 279 days (e.g., VN case). Note that the 279-day path is for a patient who returned to the hospital 272 days after discharge. Also, on average, the trajectory of the patients who were transferred from the normal care ward to the intensive care ward took 24 days.

\begin{figure}[H]
\centering
\includegraphics[width=12cm]{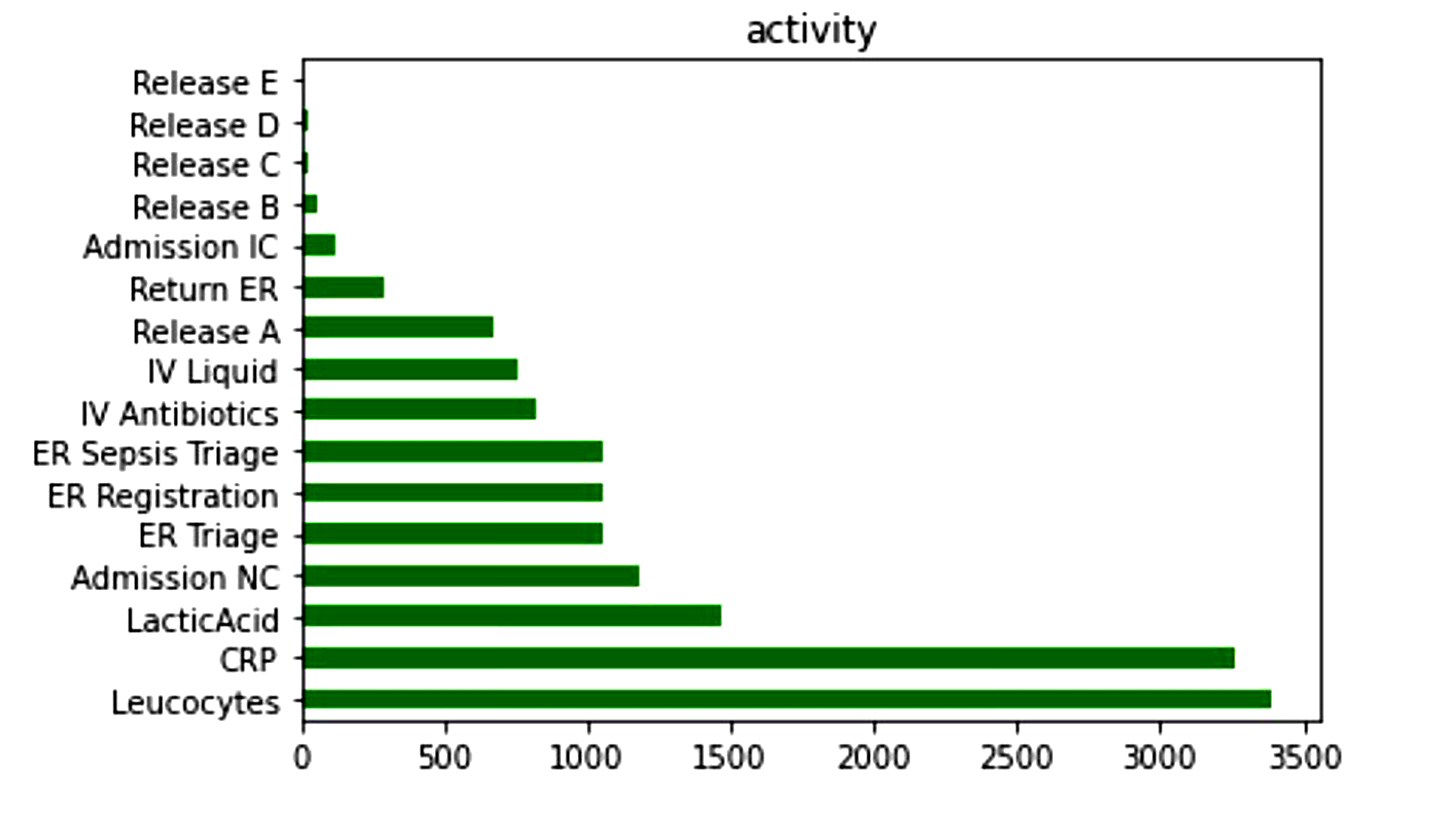}
\caption{The total number of repetitions of each activity in the event log.}
\label{fig:repitition}
\end{figure}

Moreover, a synopsis of analyzing all variants reveals that 77\% of them, i.e., 805 cases took at least a day. The measurement of CRP, Leukocytes, and Lactic Acid has been performed for 1007, 1012, and 860 cases (in 839, 843, and 752 variants). Besides, the rework amount of these indicators in the whole event log are 3262, 3383, and 1466, respectively. Hence, the condition has been such that the majority of the recorded cases were kind of required the assessment and examination of three key indicators on a regular basis. Indeed, it seems that antibiotic injection is also unavoidable for patients who have different symptoms of sepsis. In 823 cases (in 731 variants), the antibiotic injection has been accomplished, showing the importance of this activity specifically. Finally, a comparison between the number of cases transported to either normal or intensive care wards and their rework amount displays that 800 cases (in 748 variants) were sent to normal care wards and 110 cases (in 110 variants) were sent to intensive care wards so that each of which has been repeated a total of 1182 and 117 times, respectively. This indicates that a relatively small number of patients were in quite serious conditions and directed to intensive care. Based on interpretations, the total number of repetitions of activities in the event log is illustrated in Figure~\ref{fig:repitition}.

\section{Results} \label{sec:result}
We have implemented three process mining approaches on this event log. First of all, we began with HM as a well-known process discovery approach. HM parameters were set as follows:
\begin{itemize}
    \item The dependency threshold = 0.95 to discover only very strong dependencies;
    \item The long-distance threshold = 0.98 to consider potential long-distance dependencies
\end{itemize}

\begin{figure}[H]
\centering
\includegraphics[width=14cm]{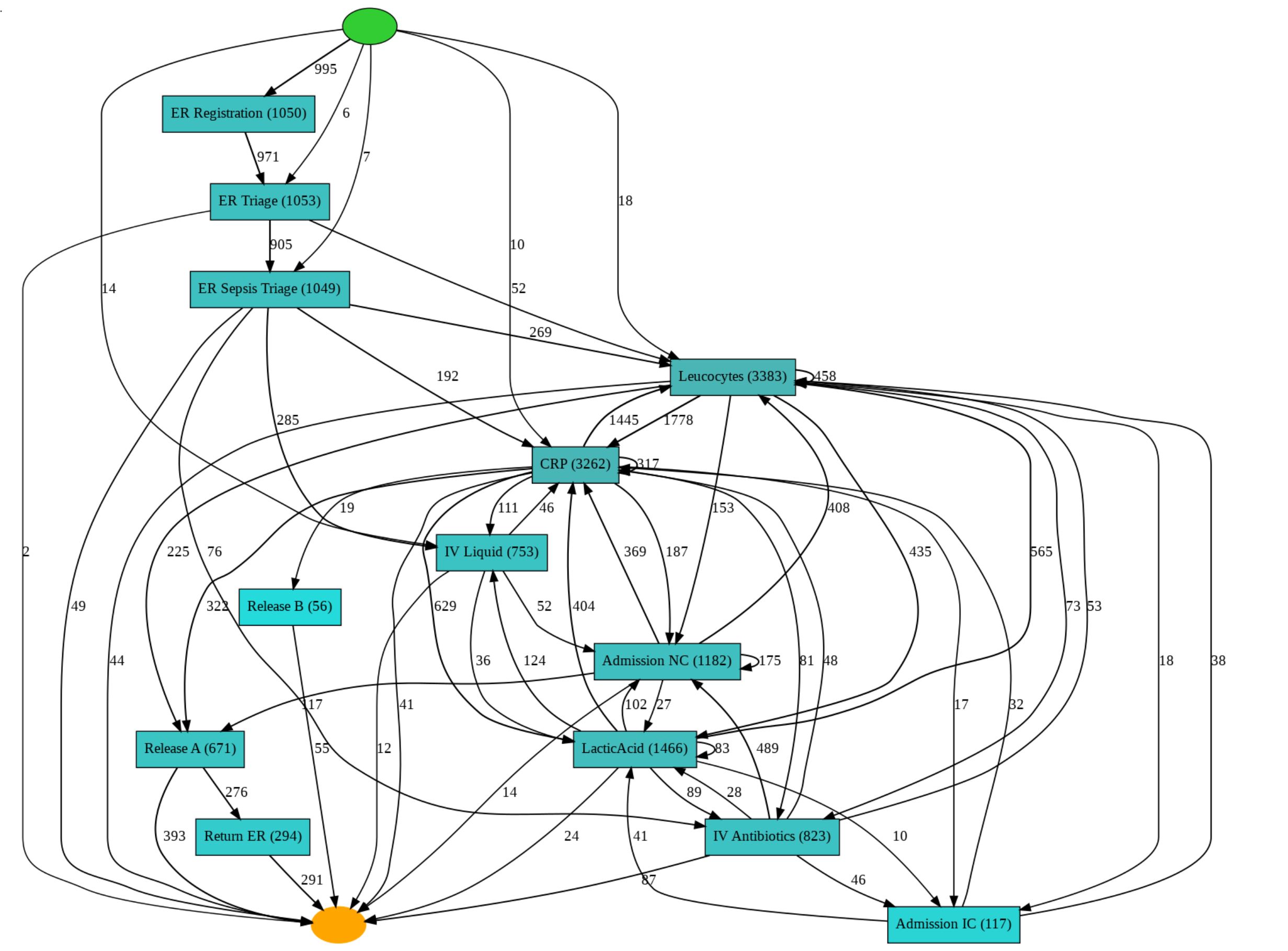}
\caption{Process model discovered by the HM.}
\label{fig:result-HM}
\end{figure}

Figure~\ref{fig:result-HM} displays the model discovered through HM. 

As it is known in Figure~\ref{fig:result-HM}, notwithstanding such settings and emphasis on strong relationships, there are too many connections to achieve a reasonable analysis of the trajectory of sepsis patients. In general, the relationships are messy and sloppy, which made it almost challenging to identify the trajectory of sepsis patients. In fact, it is extremely difficult to identify what is happening in reality from this model and match them with what is intended. HM has uncovered some of the structures that were anticipated, such as the process starting with patient registration (emergency room registration) and concluding with various discharge outcomes (such as Release A-B). Nevertheless, most of the events that occur between these registration and discharge stages are inadequately represented in the model.

Based on the result of using the multi-perspective conformance checking feature of HM, the activities that are contained in the HM model can approximately be aligned well to this event log. However, there is a substantial number of model moves recognized for activity Release A, revealing that it does not invariably pursue the activity Admission NC. Indeed, compared to the process models (i.e., the IM model and systematic process model), all other Release activities except Release A and B are missing in the HM model. This is because our focus in the process discovery was solely on the primary flow, incorporating infrequent behavior that can be characterized by deterministic rules based on the process data. The HM process model is of an average fitness of 77.4\%, a simplicity of 47\%, a generalization of 88.1\%, and an average precision of 52\% but it is not able to properly illustrate the process structure.

\begin{figure}[H]
\centering
\includegraphics[width=17cm]{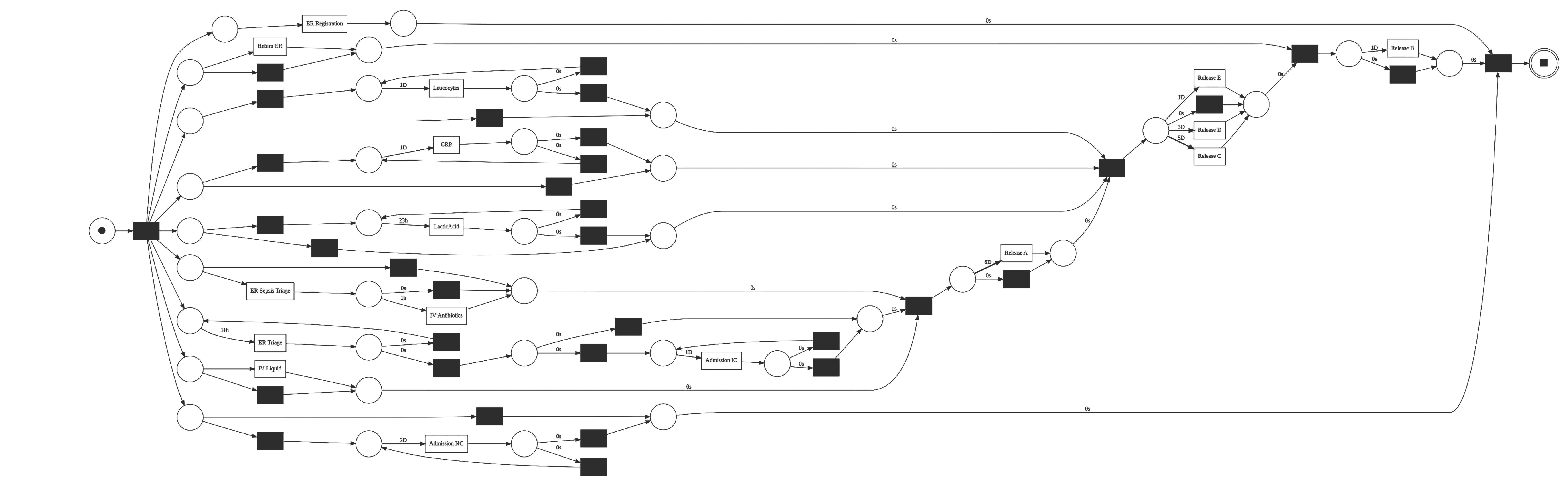}
\caption{Process model discovered by the IM.}
\label{fig:result-IM}
\end{figure}

Now, let us address the IM process model. Figure~\ref{fig:result-IM} shows The IM petri net process model. As it is clear, the IM process model is more clear than the HM process model, namely, the flows are transparent and there is no such complexity of flows here. This is an advantage of the IM compared to the HM because IM relies on the discovery of different cuts on the directly-follows graph created using the event log. Hence, after separating and finding reasonable connections, it can better get the general framework of the process.

According to Figure~\ref{fig:result-IM}, almost all the information obtained from the HM process model can be observed in the IM process model as well. However, more information can be perceived from this model. For example, based on specialists’ knowledge, administering antibiotics is not recurred in the context of the treatment in the emergency ward and this is supported by the IM process model. In this model, unlike the HM process model, more sensible connections between the registration and output activities are depicted. In addition, all outputs were discovered that were not visible in the HM process model. Note that all 16 activities were discovered in the IM process model, while the HM process model discovered only 13 activities. This implies the superiority of IM. The IM process model has an average fitness of 84.8\%, a simplicity of 62.2\%, a generalization of 90.2\%, and an average precision of 25.7\% but it is not capable of properly reflecting the structure of the process.

\begin{figure}[H]
\centering
\includegraphics[width=17cm]{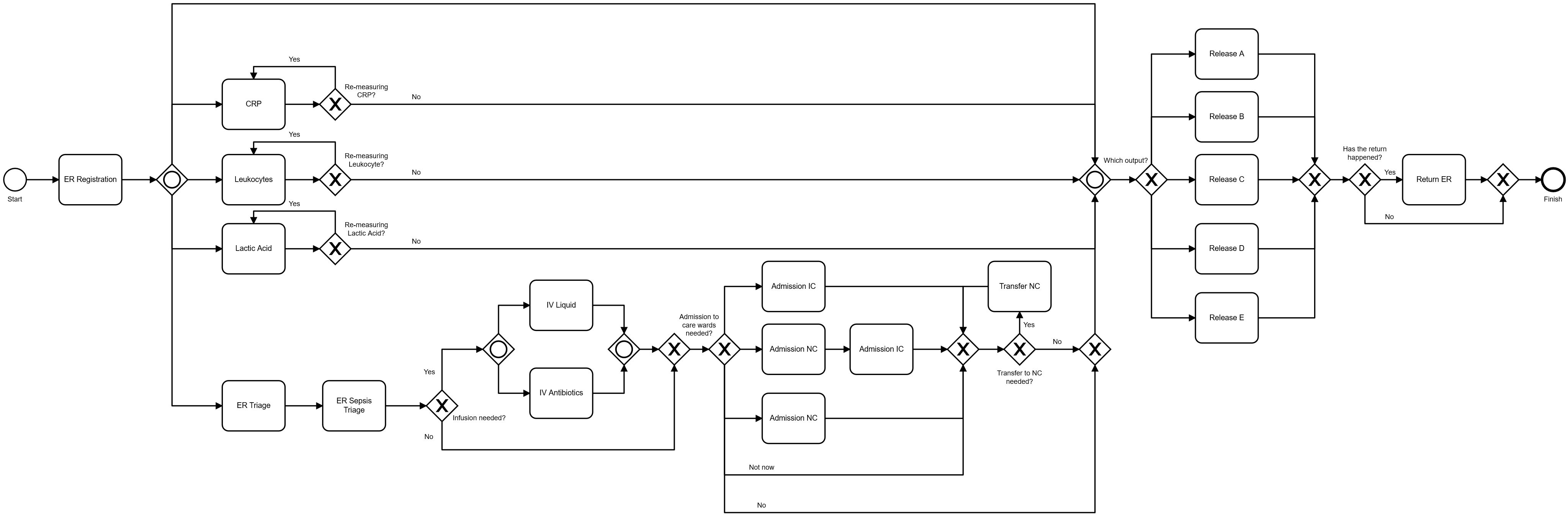}
\caption{Systematic process model by BPMN.}
\label{fig:result-bpmn}
\end{figure}

Aside from HM and IM models for the discovery of the trajectory of patients in which the results were not so satisfactory and efficient, a systematic process model is created based on the knowledge of specialists that can be used to visualize the patient trajectories explicitly and conformance checking of some medical guidelines \cite{mannhardt2017analyzing}. We acquired three patterns with respect to the information on the organizational perspective that is part of the event log. Based on the responsibility of executing each activity, we extracted three sub-logs. Specifically, each log includes all activities executed by members of its certain department. Afterward, we discovered a process model for each of the sub-logs using the IM. The systematic process model is presented in Figure~\ref{fig:result-bpmn}.

To interpret Figure~\ref{fig:result-bpmn}, results of blood tests are acquired during the entire process (i.e., Lab). First, patients are in the emergency ward (i.e., ER). In what follows, they are either admitted to a hospital ward (i.e., Admission) or depart the hospital (i.e., Discharge). Patients who are admitted are conceivably transferred (i.e., Transfer) to another hospital ward and ultimately discharged (i.e., Discharge). Finally, patients are likely to return to the emergency ward subsequently (i.e., Return ER). 

We validated the quality of the systematic process model by measuring an average fitness of 97.8\%, a simplicity of 77.7\%, a generalization of 80.21\%, and an average precision of 80.4\% of the systematic process model given to the main event log. The systematic process model conforms to most of the behavior observed in the event log. Therefore, we can use this model to reliably respond to the last three posed questions. Based on the average precision score, the models seem to be inaccurate. However, note that virtually all of the imprecision derives from the laboratory activities of CRP, Leukocytes, and Lactic Acid. Such activities may be performed in parallel to the rest of the process. When focusing on the remaining activities and abstracting from the actual behavior of the Lab, the model is more rigorously apprehended.

Almost 19\% of patients were discharged from the hospital without any treatment. We specified that 38, i.e., 3.6\% of the admitted patients are of the problematic category: They were first admitted to the normal care ward, then, transferred to the intensive care ward. About 70.2\% of the patients, i.e., 738 were admitted to the normal care ward and 6.85\% of patients, i.e., 72 were admitted to the intensive care ward. Furthermore, 56.5\% of patients return to the emergency ward within one year. Among the other patients, approximately 28\%, i.e., 294 patients returned within 28 days. Among the patients who returned to the hospital in less than 28 days, 277 were returned through A, 6 through C, 10 through D, and 1 through E. Therefore, patients of the problematic category frequently return. Consequently, the hospital could meticulously monitor these patients.

\begin{figure}[H]
\centering
\includegraphics[width=17cm]{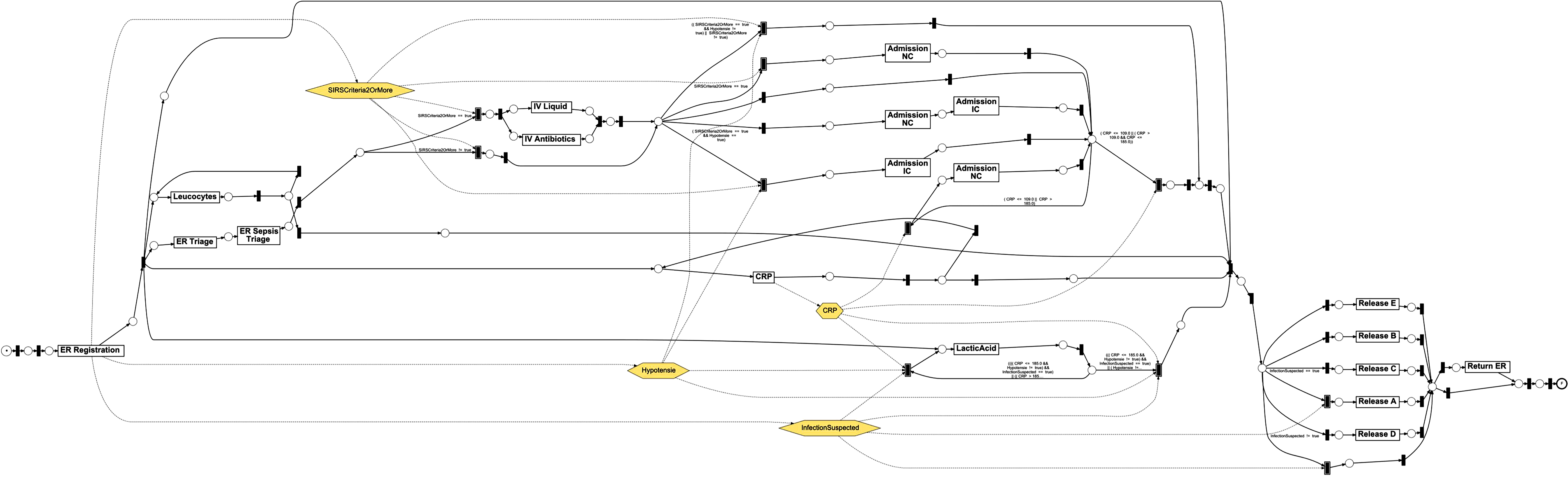}
\caption{Some decision rules discovered by the systematic process model.}
\label{fig:result-rule}
\end{figure}

We also used this model to scrutinize whether antibiotics and liquid infusions are given to sepsis patients. There are some criteria that are inspected to characterize sepsis. The event log contains the attribute SIRS Criteria 2 or More, which demonstrates whether two or more of these criteria are met. We kept merely cases for which SIRS Criteria 2 or More is true and considered such cases on the systematic process model. This divulged that 95.3\% of patients with two or more SIRS criteria finally receive antibiotics. However, based on the event log, 15\% of those patients do not get an infusion of a liquid. To survey the postponement between filling in the triage form and the infusion activities, we projected the average time between activities in the whole event log on the model. This disclosed that it takes on average 1.77 hours until antibiotics are administered. This medical guideline is violated in almost 58\% of the cases. Based on the medical guideline about measuring lactic acid within 3 hours, this is violated in almost 1.2\% of the cases as well. Such contradictions can be derived from the following reasons:
\begin{itemize}
    \item On the one hand, some guidelines suggest prescribing antibiotics within one hour and measuring lactic acid within three hours; nevertheless, it is specified that this is not constantly achievable. On the other hand, not all patients in the event log reveal adequate symptoms of sepsis. Therefore, the one-hour rule for administering antibiotics and the three-hour rule for measuring acid lactic can be kind of stringent.
    \item There can be a problem with data recording owing to the fact that the related data was documented manually by nurses and doctors. For example, we discovered some cases in which the entered antibiotics timestamp is 24 hours after the triage or other cases in which the antibiotics timestamp was before filling the triage document. Besides, in some cases, the measurement of lactic acid is after three hours, or in other cases before it. Accordingly, it is complicated to analyze these measurements as some of the diagnostics may stem from poor data quality.
\end{itemize}

we discovered several decision rules by the systematic model which is depicted in Figure~\ref{fig:result-rule}. We realized that given to the attribute SIRS criteria 2 or more, patients receive infusions. Note that patients with more than two criteria of sepsis must absolutely get infusions. We also found out decision rules in respect of the three variants of admission to the normal care and intensive care. For example, those patients who had two or more SIRS criteria and also had hypotensive were transferred to intensive care after normal care. Besides, if patients had two or more of the SIRS criteria but did not have hypotensive or had less than two SIRS criteria, they would be discharged from the hospital without the need for either normal or intensive care. Based on another discovered rule, if CRP for patients is between 109 and 185, they could leave the hospital, namely, if CRP is less than 109 or more than 185, their condition has been diagnosed as abnormal; hence, they were transferred to the admission NC ward.

To sum up, we found that the models obtained by HM and IM could not well discover the relevant process. Although it is possible to guess some cases about them, this is hardly possible. Besides, due to their complexity and vagueness, it is not possible to properly conform what is mentioned in the instructions with what is happening in order to identify potential problems. However, by using the systematic process model provided based on the knowledge of specialists and organizational information, we can get reasonable answers to the posed questions. Indeed, by corresponding the process discovery to activity patterns, it is possible to detect a model that is perceptible to specialists. Then, by imagining the patient trajectories on such a model, it is possible to share the acquired diagnostics with the process stakeholders.

\section{Discussion and managerial insight} \label{sec:discussoin}
Process mining is a group of approaches for the analysis of operational processes based on event logs extracted from an organization’s databases and information systems (e.g., ERP). Process mining brings considerable advantages to organizations. Eliminating unnecessary steps, improving performance management, and identifying and resolving process bottlenecks are a small part of its benefits. The ever-increasing importance of process mining in healthcare systems indicates the priority of managers to recognize such types of processes in order to improve their performance. Process implementation data are a worthwhile information source to amplify the management and enhancement of healthcare processes. Since the event log reflects how a process has been performed in reality, it can support specialists and healthcare organization managers with exhaustive process-related questions in the medical field. In this study, we have analyzed the process related to sepsis patients' trajectories and would like to give some recommendations/implications to the relevant managers.

The first recommendation we have for managers and stakeholders is that if they are looking to figure out the process of healthcare systems, regardless of the fact that there are various methods to discover the process, they should try to develop a process model based on organizational information and specialists' knowledge. Similar to what inductive miner does in practice, they can classify the relationship between different activities and how they are performed, then merge them to get the final process model. Given that the process structure in hospitals is usually different from each other, this makes the process structure reflect accurately and in detail for managers of each part. Through this, they get the ability to gain more insight into process flows and interactions; thus, if there are any problems, they can easily identify them.

Since there is a possibility of patients returning to the hospital, managers are advised to have more supervision of the return trajectory of patients. In fact, patients returning to the hospital results in exorbitant costs and spending additional time. Therefore, it is better to find out why they returned to the hospital. Are there any decision rules for the return of patients based on their existing symptoms and physical condition during the first visit? Determining these rules make it possible for stakeholders to identify whether there is a defect in the therapeutic or laboratory activities that leads to the return of patients to the hospital or not. Besides, if the return of patients is inevitable due to the type of their disease and the exclusive symptoms they have and has nothing to do with the structure of the process, managers can have a reasonable assessment of which patients and when they will return. Hence, they can consider measures to prevent the occurrence of bottlenecks in departments that many patients often visit and reduce waiting time in the hospital.

Acquiring data and analyzing the process should be frequently performed. Data quality issues and further questions may ensue after some initial data has been gathered. Those responsible for executing a process are advised to seek to take regular feedback from stakeholders to confirm assumptions. Iterative analysis and data enrichment can help to accurately analyze the process. On contrary, problems in recording data and poor data quality can lead managers to misperception what is really happening in the process.

\textcolor{blue}{The insights gained from process mining in this study can have several important implications for healthcare services. By analyzing the patient flow in the hospital, stakeholders can identify areas for improvement in clinical performance, particularly in relation to compliance with medical guidelines. This can lead to improvements in patient outcomes, such as reduced hospital stays, improved quality of care, and lower mortality rates. Process mining can also help to optimize resource allocation in healthcare services. By identifying bottlenecks in the patient flow, stakeholders can allocate resources more effectively, such as by adding staff to departments that are struggling with patient volume, or by streamlining processes that are causing delays or inefficiencies. This can result in cost savings for healthcare providers, as well as a better patient experience.}

\textcolor{blue}{In addition, the study highlights the importance of incorporating expert knowledge and organizational information in the development of effective process models. By combining data-driven insights with expert knowledge and organizational information, stakeholders can develop more accurate and effective process models that better reflect the realities of clinical practice. This can lead to improved decision-making, as well as more effective interventions to improve patient outcomes. The implications of this study suggest that process mining can play a valuable role in improving the quality and efficiency of healthcare services and emphasize the incorporation of expert knowledge and organizational information in order to develop an effective process model.}

\section{Conclusion} \label{sec:conclusion}
Nowadays, organizations are systematizing their operations to drive digital transformation. This is incredibly crucial in the field of healthcare. Since the health of people is considered in healthcare systems, its applications can be used to accurately identify the problems in the relevant process that leads to stakeholders being capable of providing efficient solutions. Therefore, in this study, we endeavored to discover the trajectory of sepsis patients in a Dutch hospital. After analysis of the event log and exclusively its variants, we used two well-known approaches (i.e., Heuristics and Inductive miner) to discover the relevant process, then a systematic process model was developed that emphasized the use of organizational information and specialists' knowledge. Finally, to answer the research questions, the required conformance checking was accomplished, and consequently, some decision rules were identified.

In general, we found that process mining can be exploited to elucidate and picture the ﬂow of processes in a hospital. Nonetheless, we demonstrate that in the case of concentrating on a distinctive group of patients (i.e., patients with a similar disease) and a combination of medical and logistical activities, the ﬂexibility of healthcare processes can be eluded. Besides, hospitals continually keep their processes under surveillance by exclusively examining quality indicators. However, process mining can be utilized to survey the conformance to medical guidelines in the context of therapy and logistics processes. Note that sometimes the medical guidelines are somewhat strict, and according to the patient's symptoms, it may not be necessary to follow them. Based on another key finding, known process discovery techniques are not necessarily suitable for process discovery; thus, they may illustrate models of empirical behavior that are inappropriate to be in touch with stakeholders and respond to questions. Namely, the detected process model might properly depict the practical behavior but its structure may not be discernible to the members of the process. Therefore, creating a model based on organizational information and specialists' knowledge can be beneficial and practicable for portraying the interactions between the process parties. In this regard, we used HM and IM to discover the trajectory of sepsis patients, but both of which could not lead to a proper understanding of the relevant process. However, by using the organizational information and knowledge, a systematic model was created so that not only is it quite tangible in terms of the structure but it can also be used well to answer the research questions. Finally, some decision rules were discussed in this research. Discovering such rules can lead to perspicuity on what kind of patients track a particular trajectory. It is admittedly worthwhile for the hospital to be aware of the most likely trajectory of a patient to preclude problems from stemming early in the process.

This study can be extended as follows: Future studies can obtain more extensive decision rules and perform more analysis on the trajectory taken by patients based on these rules. Besides, some patients' condition has worsened after being admitted. Given that the hospital would rather minimize the number of such patients, other studies can extensively scrutinize the influences of this class of patients on the function of the hospital in terms of the result and the time they are in the hospital. 

\newpage
\bibliographystyle{elsarticle-num}
\bibliography{cas-refs}

\end{document}